\documentclass[11pt,titlepage,a4paper]{article}
\usepackage{bozhomac}	
\usepackage{bozhlogo}	
\usepackage{cite}	
\usepackage{tabularx}	
\usepackage{varioref}	
\usepackage{xr}
\title{\bfseries	\vspace*{-2.13in}\enlargethispage{2ex}
{\huge Normal frames and linear transports\\[1ex]
	along paths in line bundles.\\[1ex]
	Applications to classical electrodynamics} 
\vspace{0.22ex} \\	\vspace{1.1ex}	{\LARGE     } }

\vspace{0.1ex}

\author{
Bozhidar Z. Iliev
\thanks{Laboratory of Mathematical Modeling in Physics,
Institute for Nuclear Research and \mbox{Nuclear} Energy,
Bulgarian Academy of Sciences,
Boul. Tzarigradsko chauss\'ee~72, 1784 Sofia, Bulgaria}
\thanks{E-mail address: bozho@inrne.bas.bg}
\thanks{URL: http://theo.inrne.bas.bg/$\sim$bozho/}
}

\date{
 \vspace{0.5ex}%
      \ShortTitle{Normal frames and transports in line bundles. Electrodynamics
								}\\[0.27ex]
 \vspace{3.2ex}
	\begin{tabular}{r@{$\colon\to~$}l}
 \vspace{0.09ex} Last updated	& March 9, 2006 	\\[0.09ex]
 \vspace{0.27ex} Produced	& \fbox{\today}	\\[0.27ex]
	\end{tabular} \\[1.27ex]
\small
	\begin{tabular}{r@{$\colon~$}l}
 \normalsize\sffamily\bfseries
  \vspace{0.27ex} http://arXiv.org e-Print archive No. &
 \normalsize\sffamily\bfseries
 math-ph/0603002
								\\[0.27ex]
	\end{tabular} \\[--0.07ex]
\normalsize
 \vspace{2.ex}{\Huge\BOZHO}	\\[2.ex]
\vspace{0.27ex}\Subject{Differential geometry, Classical electrodynamics}
\\[2.27ex]
	\begin{tabular}{r@{\hspace{0.512em}}|@{\hspace{0.512em}}l}
\vspace{0.27ex}\MSC[2000]{53B99, 53C99, 53Z05\\55R25, 83D05}
&
\vspace{0.27ex}\PACS[2003]{02.40.Vh, 11.15-q\\ 04.50.+h, 04.90.+e}
	\end{tabular} \\[1.27ex]
\vspace{0.27ex}\KeyWords{Normal frames, Frame fields\\
 			Linear transports along paths, Line bundles\\
			Inertial frames, Classical electrodynamics
			}\\[0.27ex] }

\begin{document}		

\renewcommand{\thepage}{\roman{page}}

\renewcommand{\thefootnote}{\fnsymbol{footnote}} 
\maketitle				
\renewcommand{\thefootnote}{\arabic{footnote}}   

\tableofcontents		



\begin{abstract}

The definitions and some basic properties of the linear transports along paths
in vector bundles and the normal frames for them are recalled. The formalism
is specified on line bundles and applied to a geometrical description of the
classical electrodynamics. The inertial frames for this theory are discussed.

\end{abstract}

\renewcommand{\thepage}{\arabic{page}}

\section{Introduction}
\label{Introduction}

	The transports along paths in vector bundles~\cite{bp-NF-LTP} are one
of the possible generalizations of the parallel transports in these bundles.
They are a useful tool for a geometric formulation of quantum
mechanics~\cite{bp-BQM-full}. The frames normal for them are defined as ones in
which the transports' matrices are the identity matrix; examples
of such frames are the the frames (and possibly coordinates) normal for linear
connections on vector bundles~\cite{bp-NF-D+EP}. The significance of the normal
frames (and coordinates) for the physics is a result of the assertion that they
are the mathematical concept representing the physical notion of an `inertial
frame of reference'~\cite{bp-PE-P?,bp-NF-D+EP}. From here it follows that the
(strong) equivalence principle in gravity physics is a provable
theorem~\cite{bp-PE-P?} and that the scope of its validity can be enlarged to
include the gauge theories~\cite{bp-NF-D+EP}; in particular, this is valid with
respect to classical electrodynamics~\cite{bp-EPinED}.

	The present paper contains a partial review of the general theory of
linear transports along paths in vector bundles and the frames normal for them.
It is exemplified on 1\ndash dimensional vector bundles, known as line bundles.
The formalism is then applied to a geometric description of the classical
electrodynamics and the inertial frames for it.

	Sections~\ref{Sect2} and~\ref{Sect3} contain the definitions and some
basic properties of the linear transports along paths in vector bundles and of
the frames normal for them, respectively. The proofs, extended versions of
these results and a lot of details on that items can be found
in~\cite{bp-NF-LTP}. Section~\ref{4-Sect5n} specifies the results, concerning
linear transports and frames normal for them, on line bundles.

	In section~\ref{Sect5}, the results obtained are applied to a
geometric description of the classical electromagnetic field.
Section~\ref{Sect6} is devoted to a brief discussion of the inertial frames for
the classical electromagnetic field. Section~\ref{Conclusion} closes the paper.

\section{Linear transports along paths (brief review)}
\label{Sect2}

	Let $(E,\pi,B)$ be a complex~%
\footnote{~%
All of our definitions and results hold also for real vector bundles. Most of
them are valid for vector bundles over more general fields too but this is
inessential for the following.%
}
vector bundle~\cite{Poor,Greub&et_al.-1} with bundle (total) space $E$, base
$B$, projection $\pi\colon E\to B$, and homeomorphic fibres
$\pi^{-1}(x)$, $x\in B$.~%
\footnote{~%
When writing $x\in X$, $X$ being a set, we mean ``for all $x$ in $X$'' if
the point $x$ is not specified (fixed, given) and is considered as an
argument or a variable.%
}
The base $B$ is supposed to be  a $C^1$ differentiable manifold.
By $J$ and $\gamma\colon J\to B$ are denoted real interval  and path in $B$,
respectively. The paths considered are generally \emph{not} supposed to be
continuous or differentiable unless their differentiability class is stated
explicitly. If $\gamma$ is a $C^1$ path, the vector field tangent to it is
denoted by $\dot{\gamma}$.

	\begin{Defn}	\label{4-Defn2.1}
	\index{linear transport along paths!definition|defined}
	A \emph{linear transport along paths} in the bundle $(E,\pi,B)$ is a
map $L$ assigning to every path $\gamma$ a map $L^\gamma$,
\emph{transport along} $\gamma$, such that
$L^\gamma\colon (s,t)\mapsto L^\gamma_{s\to t}$ where the map
	\begin{equation}	\label{4-2.1}
L^\gamma_{s\to t} \colon  \pi^{-1}(\gamma(s)) \to \pi^{-1}(\gamma(t))
	\qquad s,t\in J,
	\end{equation}
called \emph{transport along $\gamma$ from $s$ to} $t$, has the properties:
	\begin{alignat}{2}	\label{4-2.2}
L^\gamma_{s\to t}\circ L^\gamma_{r\to s} &=
			L^\gamma_{r\to t},&\qquad  r,s,t&\in J, \\
L^\gamma_{s\to s} &= \id_{\pi^{-1}(\gamma(s))}, & s&\in J,	\label{4-2.3}
\\
L^\gamma_{s\to t}(\lambda u + \mu v) 				\label{4-2.4}
  &= \lambda L^\gamma_{s\to t}u + \mu L^\gamma_{s\to t}v,
	& \lambda,\mu &\in \mathbb{C},\quad u,v\in{\pi^{-1}(\gamma(s))},
	\end{alignat}
where  $\circ$ denotes composition of maps and $\id_X$ is the identity map of
a set $X$.
	\end{Defn}

	Let $\{e_i(s;\gamma)\}$ be a $C^1$ basis in $\pi^{-1}(\gamma(s))$,
$s\in J$.~%
\footnote{~%
Here and henceforth the Latin indices run from 1 to
$\dim \pi^{-1}(x),\ x\in B$. We also assume the Einstein summation rule on
indices repeated on different levels.%
}
So, along $\gamma\colon J\to B$ we have a set $\{e_i\}$ of bases on
$\pi^{-1}(\gamma(J))$ such that the liftings $\gamma\mapsto e_i(\cdot,\gamma)$
of paths are of class $C^1$.

	The \emph{matrix}
 $\Mat{L}(t,s;\gamma):=\bigl[\Sprindex[L]{j}{i}(t,s;\gamma)\bigr]$
\emph{(along $\gamma$ at $(s,t)$ in $\{e_i\}$) of a linear transport} $L$
along $\gamma$ from $s$ to $t$ is defined via the expansion~%
\footnote{~%
Notice the different positions of the arguments $s$ and $t$ in
$L_{s\to t}^{\gamma}$ and in $\Mat{L}(t,s;\gamma)$.%
}
\(
L_{s\to t}^{\gamma} \bigl(e_i(s;\gamma)\bigr)
		=:\Sprindex[L]{i}{j}(t,s;\gamma) e_j(t;\gamma)
		\ s,t\in J.
\)
A change
\(
  \{ e_i(s;\gamma) \} \mapsto
	\{ e_i^\prime(s;\gamma)	:= A_{i}^{j}(s;\gamma)e_j(s;\gamma) \}
\)
via of a non\ndash degenerate matrix
$A(s;\gamma):=\bigl[A_{i}^{j}(s;\gamma)\bigr]$ implies
	\begin{gather}	\label{4-2.10}
\Mat{L}(t,s;\gamma)\mapsto\Mat{L}^\prime(t,s;\gamma)
  = A^{-1}(t;\gamma) \Mat{L}(t,s;\gamma) A(s;\gamma)
	\end{gather}
or in component form
\(
\Sprindex[L]{i}{\prime\mspace{0.92mu} j}(t,s;\gamma)
   = \bigl(A^{-1}(t;\gamma)\bigr)_{k}^{j}
     \Sprindex[L]{l}{k}(t,s;\gamma) A_{i}^{l}(s;\gamma).
\)

	\begin{Prop}	\label{4-Prop2.4}
	A non\ndash degenerate matrix\ndash valued function
$\Mat{L}\colon (t,s;\gamma)\mapsto\Mat{L}(t,s;\gamma)$
is a matrix of some linear transport along paths $L$ (in a given field
$\{e_i\}$ of bases along $\gamma$) iff
	\begin{equation}	\label{4-2.14}
\Mat{L}(t,s;\gamma) = \Mat{F}^{-1}(t;\gamma) \Mat{F}(s;\gamma)
	\end{equation}
where $\Mat{F}\colon (t;\gamma)\mapsto\Mat{F}(t;\gamma)$ is a non\ndash degenerate
matrix\nobreakdash-valued function.
	\end{Prop}

	\begin{Prop}	\label{4-Prop2.5}
	If the matrix $\Mat{L}$ of a linear transport $L$ along paths has a
representation
\(
\Mat{L}(t,s;\gamma)
   = \lindex[\mspace{-2mu}\Mat{F}]{}{\star}^{-1}(t;\gamma)
     \lindex[\mspace{-2mu}\Mat{F}]{}{\star}(s;\gamma)
\)
for some matrix-valued function
$\lindex[\mspace{-2mu}\Mat{F}]{}{\star}(s;\gamma)$,
then all matrix\nobreakdash-valued functions $\Mat{F}$ representing $\Mat{L}$
via~\eref{4-2.14} are given by
\(
\Mat{F}(s;\gamma) = \Mat{D}^{-1}(\gamma)
   \lindex[\mspace{-2mu}\Mat{F}]{}{\star}(s;\gamma)
\)
where $\Mat{D}(\gamma)$ is a non\ndash degenerate matrix depending only on $\gamma$.
	\end{Prop}

	Let $\{e_i(s;\gamma)\}$ be a smooth field of bases along
$\gamma\colon J\to B,\ s\in J$. The explicit local action of the derivation
$D\colon\gamma\mapsto D^\gamma \colon s\mapsto D_s^\gamma$, associated to $L$,
on a $C^1$ lifting of paths $\lambda$ is
	\begin{equation}	\label{4-2.22}
D_s^\gamma\lambda
   = \biggl[ \frac{\od \lambda_\gamma^i(s)}{\od s}
     + \Sprindex[\Gamma]{j}{i}(s;\gamma)\lambda_\gamma^j(s)
     \biggr]  e_i(s;\gamma) .
	\end{equation}
Here the \emph{(2-index) coefficients}
$\Sprindex[\Gamma]{j}{i}$ of the linear transport $L$ are defined by
	\begin{equation}	\label{4-2.23}
\Sprindex[\Gamma]{j}{i}(s;\gamma)
   := \frac{\pd\Sprindex[L]{j}{i}(s,t;\gamma)}{\pd t}\bigg|_{t=s}
   = - \frac{\pd\Sprindex[L]{j}{i}(s,t;\gamma)}{\pd s}\bigg|_{t=s}
	\end{equation}
and, evidently, uniquely determine the derivation $D$ generated by $L$.
If a matrix $\Mat{F}$ determines the matrix $\Mat{L}$ of a transport $L$
according to proposition~\ref{4-Prop2.4}, then
	\begin{equation}	\label{4-2.25}
\Mat{\Gamma}(s;\gamma):=\bigl[ \Sprindex[\Gamma]{j}{i}(s;\gamma) \bigr]
   = \genfrac{.}{|}{}{}{\pd\Mat{L}(s,t;\gamma)}{\pd t}_{t=s}
   = \Mat{F}^{-1}(s;\gamma)\frac{\od \Mat{F}(s;\gamma)}{\od s} .
	\end{equation}

A change
$\{e_i\}\to\{e_{i}^{\prime}=A_{i}^{j}e_i\}$ of the bases along a path $\gamma$
with a non\ndash degenerate $C^1$ matrix\ndash valued function
$A(s;\gamma):=\bigl[A_{i}^{j}(s;\gamma)\bigr]$ implies
\(
\Mat{\Gamma}(s;\gamma) = \bigl[ \Sprindex[\Gamma]{j}{i}(s;\gamma) \bigr]
	\mapsto  \Mat{\Gamma}^\prime(s;\gamma)
   = \bigl[ \Sprindex[\Gamma]{j}{\prime\,i}(s;\gamma) \bigr]
\)
with
	\begin{equation}	\label{4-2.26}
\Mat{\Gamma}^\prime(s;\gamma)
  = A^{-1}(s;\gamma) \Mat{\Gamma}(s;\gamma) A(s;\gamma)
    + A^{-1}(s;\gamma)\frac{\od A(s;\gamma)}{\od s}.
	\end{equation}

\section
[Normal frames for linear transports (definitions and some results)]
{Normal frames for linear transports\\ (definitions and some results)}
\label{Sect3}

	Let a linear transport $L$ along paths be given in a vector bundle
$(E,\pi,B)$, $U\subseteq B$ be an arbitrary subset in $B$, and
$\gamma\colon J\to U$ be a path in $U$.

	\begin{Defn}	\label{4-Defn3.1}
\index{normal frame!in vector bundles!basic definitions|(}
	A \emph{frame field (of bases)} in $\pi^{-1}(\gamma(J))$ is called
\emph{normal along $\gamma$ for} $L$ if the matrix of $L$ in it is the
identity matrix along the given path $\gamma$.
%
	A \emph{frame field (of bases)} defined on $U$ is
called \emph{normal on $U$ for} $L$ if it is normal along every path
$\gamma\colon J\to U$ in $U$. The \emph{frame} is called \emph{normal for}
$L$ if $U=B$.
\index{normal frame!in vector bundles!basic definitions|)}
	\end{Defn}

	\begin{Defn}	\label{4-Defn3.3}
\index{linear transport along paths!Euclidean!basic definitions|(}
	A \emph{linear transport along paths} (or \emph{along a path $\gamma$})
is called \emph{Euclidean along some (or the given) path} $\gamma$ if it admits
a frame normal along $\gamma$.
%
	A \emph{linear transport along paths} is called \emph{Euclidean} on
$U$ if it admits frame(s) normal on $U$. It is called \emph{Euclidean} if
$U=B$.
\index{linear transport along paths!Euclidean!basic definitions|)}
	\end{Defn}

	\begin{Prop}	\label{4-Prop3.1}
	The following statements are equivalent in a given frame $\{e_i\}$
over $U\subseteq B$:

\begin{subequations}	\label{4-3.1}
\indent
\textup{\textbf{\hphantom{v}(i)}}
The matrix of $L$ is the identity matrix on $U$, \ie
\(
\Mat{L}(t,s;\gamma) = \openone
\)
along every path $\gamma$ in $U$.

\textup{\textbf{\hphantom{i}(ii)}}
The matrix of $L$ along every $\gamma\colon J\to U$
depends only on $\gamma$, \ie it is independent of the points at which it is
calculated:
\(\Mat{L}(t,s;\gamma) = C(\gamma)
\)
where $C$ is a matrix-valued function of $\gamma$.

\textup{\textbf{\hphantom{}(iii)}}
If $E$ is a $C^1$ manifold, the coefficients
$\Sprindex[\Gamma]{j}{i}(s;\gamma)$ of $L$ vanish on $U$, \ie
\(\Mat{\Gamma}(s;\gamma) = 0
\)
along every path $\gamma$ in $U$.

\textup{\textbf{\hphantom{}(iv)}}
The explicit local action of the derivation $D$ along paths generated by $L$
reduces on $U$ to differentiation of the components of the liftings with
respect to the path's parameter if the path lies entirely in $U$:
\(
D_{s}^{\gamma}\lambda
		= \frac{\od\lambda_\gamma^i(s)}{\od s} \, e_i(s;\gamma)
\)
where $\lambda=\lambda^ie_i$
is a $C^1$ lifting of paths and
$\lambda\colon\gamma\mapsto\lambda_\gamma$.

\textup{\textbf{\hphantom{i}(v)}}
The transport $L$ leaves the vectors' components
unchanged along any path in $U$, \viz we have
\(
L_{s\to t}^{\gamma}\bigl(u^i e_i(s;\gamma)\bigr) = u^i e_i(t;\gamma)
\)
for all $u^i\in\mathbb{C}$.

\textup{\textbf{\hphantom{}(vi)}}
The basic vector fields are $L$\ndash transported
along any path $\gamma\colon J\to U$:
\(
L_{s\to t}^{\gamma} \bigl(e_i(s;\gamma)\bigr) =e_i(t;\gamma).
\)
	\end{subequations}
	\end{Prop}

	\begin{Rem}	\label{4-Rem3.1}
It is valid the equivalence
\(
\Mat{L}(t,s;\gamma) = \openone \iff \Mat{F}(s;\gamma) = \Mat{B}(\gamma)
\)
with $\Mat{B}$ being a matrix-valued function of the path $\gamma$ only.
According to proposition~\ref{4-Prop2.5}, this dependence is inessential and,
consequently, in a normal frame, we can always choose
representation~\eref{4-2.14} with
\(
\Mat{F}(s;\gamma) = \openone.
\)
	\end{Rem}

	\begin{Cor}	\label{4-Cor4.1}
	Every linear transport along paths is Euclidean along every fixed
path without self\ndash intersections.
	\end{Cor}

	\begin{Thm}	\label{4-Thm3.1}
	A linear transport along paths admits frames normal on some set
(resp.\ along a given path) if and only if its action along every path in
this set (resp.\ along the given path) depends only on the initial and final
point of the transportation but not on the particular path connecting these
points. In other words, a transport is Euclidean on $U\subseteq B$ iff it
is path\ndash independent on $U$.
	\end{Thm}


	\begin{Prop}	\label{4-Prop5.1}
Let $L$ be a linear transport along paths in $(E,\pi,M)$, $E$ and $M$ being
$C^1$ manifolds, and $L$  be Euclidean on $U\subseteq M$
(resp.\ along a $C^1$ path $\gamma\colon J\to M$). Then the matrix
$\Mat{\Gamma}$ of its coefficients has the representation
	\begin{equation}	\label{4-5.1}
\Mat{\Gamma}(s;\gamma)
   = \sum_{\mu=1}^{\dim M} \Gamma_\mu(\gamma(s)) \dot\gamma^\mu(s)
   \equiv \Gamma_\mu(\gamma(s)) \dot\gamma^\mu(s)
	\end{equation}
in any frame $\{e_i\}$ along every (resp.\ the given) $C^1$ path
$\gamma\colon J\to U$, where
\(
{\Gamma}_\mu =
  \bigl[\Sprindex[\Gamma]{j\mu}{i}\bigr]_{i,j=1}^{\dim\pi^{-1}(x)}
\)
are some matrix-valued functions, defined on an open set~$V$ containing $U$
(resp.\ $\gamma(J)$) or equal to it, and $\dot\gamma^\mu$ are the components
of $\dot\gamma$ in some frame $\{E_\mu\}$ along $\gamma$ in the bundle space
tangent to $M$, $\dot\gamma=\dot\gamma^\mu E_\mu$. The functions
$\Sprindex[\Gamma]{j\mu}{i}$ are termed \emph{3\ndash index coefficients} of
$L$.
	\end{Prop}

	Let $U$ be an open set, \eg $U=M$. If we change the frame $\{E_\mu\}$
in the bundle space tangent to $M$,
$\{E_\mu\}\mapsto \{E_\mu^{\prime} = B_{\mu}^{\nu}E_\nu\}$ with
$B=\bigl[B_{\mu}^{\nu}\bigr]$ being non\ndash degenerate matrix\ndash valued
function, and simultaneously the bases in the fibres $\pi^{-1}(x)$, $x\in M$,
$\{e_i|_x\} \mapsto \{e_i^{\prime}|_x = A_{i}^{j}(x)e_j|_x\}$, then,
from~\eref{4-2.26} and~\eref{4-5.1}, we see that ${\Gamma}_\mu$ transforms
into ${\Gamma}_\mu^{\prime}$ such that
	\begin{equation}	\label{4-5.3}
{\Gamma}_\mu^{\prime}
  = B_{\mu}^{\nu} A^{-1}{\Gamma}_\nu A  + A^{-1} E'_\mu(A)
  = B_{\mu}^{\nu} A^{-1}\bigl( {\Gamma}_\nu A  + E_\nu(A) \bigr)
	\end{equation}
where $A:=\bigl[A_{i}^{j}\bigr]_{i,j=1}^{\dim\pi^{-1}(x)}$ is
non\ndash degenerate and of class $C^1$.

	\begin{Thm}	\label{4-Thm5.1}
	A $C^2$ linear transport $L$ along paths is Euclidean on a
neighborhood $U\subseteq M$ if and only if in every frame the matrix
$\Mat{\Gamma}$ of its coefficients has a representation~\eref{4-5.1} along
every $C^1$ path $\gamma$ in $U$ in which the matrix\nobreakdash-valued
functions $\Gamma_\mu$, defined on an open set containing $U$ or equal to it,
satisfy the equalities
	\begin{equation}	\label{4-5.5}
\bigl( R_{\mu\nu}(-{\Gamma}_1,\ldots,-{\Gamma}_{\dim M}) \bigr) (x)
  = 0
	\end{equation}
where $x\in U$ and
	\begin{equation}	\label{4-5.6}
R_{\mu\nu}(-{\Gamma}_1,\ldots,-{\Gamma}_{\dim M})
  := - \frac{\pd{\Gamma}_\mu}{\pd x^\nu}
     + \frac{\pd{\Gamma}_\nu}{\pd x^\mu}
     + {\Gamma}_\mu{\Gamma}_\nu
     - {\Gamma}_\nu{\Gamma}_\mu .
\end{equation}
in a coordinate frame
$\bigl\{E_\mu=\frac{\pd}{\pd x^\mu}\bigr\}$ in a neighborhood of $x$
	\end{Thm}

	\begin{Thm}	\label{4-Thm5.2}
A linear transport $L$ along paths is Euclidean on a submanifold $N$ of $M$
if and only if in every frame $\{e_i\}$, in the bundle space over $N$, the
matrix of its coefficients has a representation~\eref{4-5.1} along every
$C^1$ path in $N$ and, for every $p_0\in N$ and a chart $(V,x)$ of  $M$
such that $V\ni p_0$ and
$x(p)=( x^1(p),\dots,x^{\dim N}(p),t_0^{\dim N+1},\dots,t_0^{\dim M} )$
for every $p\in N\cap V$ and constant numbers
$t_0^{\dim N+1},\dots,t_0^{\dim M}$, the equalities
	\begin{equation}	\label{4-5.14}
\bigl(  R^N_{\alpha\beta}(-\Gamma_1,\ldots,-\Gamma_{\dim N}) \bigr) (p) = 0,
\qquad
\alpha,\beta=1,\dots,\dim N
	\end{equation}
hold for all $p\in N\cap V$ and
	\begin{equation}	\label{4-5.15}
R^N_{\alpha\beta}(-\Gamma_1,\ldots,-\Gamma_{\dim N})
:= R_{\alpha\beta}(-\Gamma_1,\ldots,-\Gamma_{\dim M})
 =   - \frac{\pd \Gamma_\alpha} {\pd x^\beta}
     - \frac{\pd \Gamma_\beta}  {\pd x^\alpha}
     + \Gamma_\alpha \Gamma_\beta
     - \Gamma_\beta  \Gamma_\alpha .
	\end{equation}
Here $\Gamma_1,\ldots,\Gamma_{\dim N}$ are first $\dim N$ of the matrices of
the 3-index coefficients of $L$ in the coordinate frame
$\bigl\{\frac{\pd}{\pd x^\mu}\bigr\}$ in the tangent bundle space over
$N\cap V$.
	\end{Thm}

\section{Linear transports and normal frames in line bundles}
\label{4-Sect5n}

	Let $(E,\pi,M)$ be one-dimensional vector bundle over a $C^1$
manifold $M$; such bundles are called line bundles. Thus the (typical) fibre of
$(E,\pi,M)$ can be identified with $\mathbb{C}$ (resp.\ $\mathbb{R}$ in the
real case) and then the fibre $\pi^{-1}(x)$ over $x\in M$ will be an isomorphic
image of $\mathbb{C}$ (resp.\ $\mathbb{R}$ in the real case). Let $\gamma\colon
J\to M$ be of class $C^1$ and $L$ be a linear transport along paths in
$(E,\pi,M)$. A frame $\{e\}$ along $\gamma$ consists of a single non\ndash zero
vector field $e\colon (s;\gamma)\to
e(s;\gamma)\in\pi^{-1}(\gamma(s))\backslash\{0\}$, $s\in J$, and in it the
matrix of $L^\gamma$ at $(t,s)\in J\times J$ is simply a number
$\Mat{L}(t,s;\gamma)\in\mathbb{C}$,
 $L_{s\to t}^{\gamma}(u e(s;\gamma)) = u \Mat{L}(t,s;\gamma) e(t;\gamma)$
for $u\in\mathbb{C}$ and $s,t\in J$. By proposition~\ref{4-Prop2.4}, the
general form of $\Mat{L}$ is
	\begin{equation}	\label{4-5n.1}
\Mat{L}(t,s;\gamma) = \frac{f(s;\gamma)}{f(t;\gamma)}
	\end{equation}
where $f\colon (s;\gamma)\mapsto f(s;\gamma)\in\mathbb{C}\backslash\{0\}$ is
defined up to (left) multiplication with a function of $\gamma$
(proposition~\ref{4-Prop2.5}). Respectively, due to~\eref{4-2.25}, the matrix
of the coefficient(s) of $L$ is
	\begin{equation}	\label{4-5n.2}
\Mat{\Gamma}(s;\gamma)
= \frac{\pd \Mat{L}(t,s;\gamma)}{\pd s}\Big|_{t=s}
= \frac{1}{f(s;\gamma)} \frac{\od f(s;\gamma)}{\od s}
= \frac{\od}{\od s}\bigl[ \ln(f(s;\gamma)\bigr]
	\end{equation}
and~\cite[equation~\eref{LB-4-2.27}]{bp-NF-LTP} takes the form
	\begin{equation}	\label{4-5n.3}
\Mat{L}(t,s;\gamma)
= \exp \biggl(
	- \int\limits_{s}^{t} \Mat{\Gamma}(\sigma;\gamma) \Id\sigma
	\biggr).
	\end{equation}

	A change $e(s;\gamma)\mapsto e'(s;\gamma)=a(s;\gamma)e(s;\gamma)$,
with $a(s;\gamma)\in\mathbb{C}\backslash\{0\}$, of the frame $\{e\}$ implies
(see~\eref{4-2.10} and~\eref{4-2.26})
	\begin{subequations}	\label{4-5n.4}
	\begin{gather}	\label{4-5n.4a}
\Mat{L}(t,s;\gamma) \mapsto \Mat{L}'(t,s;\gamma)
	= \frac{a(s;\gamma)}{a(t;\gamma)} \Mat{L}(t,s;\gamma)
\\	\label{4-5n.4b}
\Mat{\Gamma}(s;\gamma) \mapsto \Mat{\Gamma}'(s;\gamma)
	= \Mat{\Gamma}(s;\gamma) +
		\frac{\od}{\od s}\bigl[ \ln(a(s;\gamma)\bigr] .
	\end{gather}
	\end{subequations}

	The explicit local action of the derivation $D$ along paths generated
by $L$ is
	\begin{equation}	\label{4-5n.5}
D_{s}^{\gamma} \lambda
= \Bigl(
	\frac{\od\lambda_\gamma(s)}{\od s}
	+ \Mat{\Gamma}(s;\gamma) \lambda_\gamma(s)
  \Bigr) e(s;\gamma)
	\end{equation}
where $\lambda\in\PLift^1(E,\pi,M)$ and~\eref{4-2.22} was used.

	Let us now look on the normal frames on one-dimensional vector
bundles.

	A frame $\{e\}$ is normal for $L$ along $\gamma$ (resp.\ on $U$) iff
in that frame equation~\eref{4-5n.1} holds with
	\begin{equation}	\label{4-5n.6}
f(s;\gamma) = f_0(\gamma)
	\end{equation}
where  $\gamma\colon J\to M$ (resp.\ $\gamma\colon J\to U$) and
$f_0\colon\gamma\mapsto f_0(\gamma)\in\mathbb{C}\backslash\{0\}$ (see
remark~\ref{4-Rem3.1} and proposition~\ref{4-Prop3.1}). Since, in a frame
normal along $\gamma$ (resp.\ on $U$), it is fulfilled
	\begin{equation}	\label{4-5n.7}
\Mat{L}(t,s;\gamma) = \openone,
\quad
\Mat{\Gamma}(s;\gamma) = 0
	\end{equation}
for the given path $\gamma$ (resp.\ every path in $U$),
in every frame $\{e'=ae\}$, we have
	\begin{equation}	\label{4-5n.8}
\Mat{L}'(t,s;\gamma) = \frac{a(s;\gamma)}{a(t;\gamma)},
\quad
\Mat{\Gamma}'(s;\gamma) = \frac{\od}{\od s}\bigl[ \ln(a(s;\gamma)\bigr] .
	\end{equation}
In addition, for Euclidean on $U\subseteq M$ transport $L$, the representation
	\begin{equation}	\label{4-5n.9}
\Mat{\Gamma}'(s;\gamma) = \Gamma'_\mu(\gamma(s)) \dot\gamma^{\prime\,\mu}(s)
	\end{equation}
holds for every $C^1$ path $\gamma\colon J\to U$ and some
$\Gamma'_\mu\colon V\to\mathbb{C}$ with $V$ being an open set such that
$V\supseteq U$ (proposition~\ref{4-Prop5.1}). This means (see
theorem~\ref{4-Thm3.1} and~\cite[theorem~\ref{LB-4-Thm3.2}]{bp-NF-LTP})
that~\eref{4-5n.8} holds for
	\begin{equation}	\label{4-5n.10}
a(s;\gamma)=a_0(\gamma(s)),
	\end{equation}
where $a_0\colon U\to\mathbb{C}\backslash\{0\}$, and, consequently, the
equality~\eref{4-5n.9} can be satisfied if we choose
	\begin{equation}	\label{4-5n.11}
\Gamma'_\mu = E_\mu(a)
	\end{equation}
with $a\colon V\to\mathbb{C}$, $a|_U=a_0$ and $\{E_\mu\}$ being a frame in
the bundle space tangent to $M$ which, in particular, can be a coordinate
one, $E_\mu=\frac{\pd}{\pd x^\mu}$.  Of course, if $U$ is not an open set,
this choice of $\Gamma'_\mu$ is not necessary; for
example, the equality~\eref{4-5n.9} will be preserved, if to the r.h.s.\
of~\eref{4-5n.11} is added a function $G'_\mu$ such that
$G'_\mu\dot\gamma^{\prime\,\mu}=0$.

	By virtue of~\eref{4-5.3}, the functions $\Gamma_\mu$ and
$\Gamma'_\mu$ in two arbitrary pairs of frames
$(\{e\},\{E_\mu\})$ and
$(\{e'=ae\},\{E'_\mu=B_\mu^\nu E_\nu\})$, respectively, are connected via
	\begin{equation}	\label{4-5n.12}
\Gamma'_\mu
= B_\mu^\nu \Gamma_\nu + \frac{1}{a} E'_\mu(a)
= B_\mu^\nu \bigl( \Gamma_\nu +  E_\nu(\ln a) \bigr)
	\end{equation}
and, consequently, with respect to changes of the frames in the tangent
bundle space over $M$, when $a=1$, they behave like the components of a
covariant vector field (one\ndash form).
Therefore, on an open set $U$, \eg $U=M$, the quantity
	\begin{equation}	\label{4-5n.13}
\omega = \Gamma_\mu E^\mu,
	\end{equation}
where $\{E^\mu\}$ is the coframe dual to $\{E_\mu\}$ (in local coordinates:
$E_\mu=\frac{\pd}{\pd s^\mu}$ and $E^\mu= \od x^\mu$), is a 1\ndash form
over $M$ (with respect to changes of the local coordinates on $M$ or of
the frames in the (co)tangent bundle space over $M$). However, it depends
on the choice of the frame $\{e\}$ in the bundle space $E$ and a change
$e\mapsto e'=ae$ implies
	\begin{equation}	\label{4-5n.14}
\omega\mapsto\omega'
= \omega + (E_\nu(\ln a))E^\nu
= \omega + (E'_\nu(\ln a))E^{\prime\,\nu} .
	\end{equation}

	Using the 1-form~\eref{4-5n.13}, we see that
	\begin{equation}	\label{4-5n.15}
\Mat{\Gamma}(s;\gamma)=\omega|_{\gamma(s)}(\dot\gamma(s))
	\end{equation}
and~\eref{4-5n.3} can be rewritten as
	\begin{equation}	\label{4-5n.16}
\Mat{L}(t,s;\gamma)
= \exp \biggl(
	- \int\limits_{\gamma(s)}^{\gamma(t)} \omega
	\biggr)
	\end{equation}
where the integration is along some path in $U$ (on which the transport $L$ is
Euclidean). Hence $\Mat{L}$ (or $L$) depends only on the points $\gamma(s)$
and $\gamma(t)$, not on the particular path connecting them, as it should be
(theorem~\ref{4-Thm3.1}). The self\ndash consistency of our results is
confirmed by the equation
	\begin{equation}	\label{4-5n.17}
R_{\mu\nu}|_U = 0
	\end{equation}
which is a consequence of~\eref{4-5n.11} and~\eref{4-5.6} and which is a
necessary and sufficient condition for the existence of frames normal on an
open set $U$ (theorem~\ref{4-Thm5.1}).

	We end this section with a remark that frames normal along injective
paths always exist (corollary~\ref{4-Cor4.1}), but on an arbitrary submanifold
$N\subseteq M$ they exist iff the functions $\Gamma_\mu$ satisfy the
conditions~\eref{4-5.14} with $x\in N$ in the coordinates described in
theorem~\ref{4-Thm5.2}.

\section{Bundle description of the classical electromagnetic field}
	\label{Sect5}

	Now we would like to apply the above formalism to a description of
the classical electromagnetic field. Before going on, we should say that the
accepted natural formalism in gauge field theories, in particular in the
electrodynamics, is via connections on vector
bundles~\cite{Konopleva&Popov,Drechsler&Mayer}. This approach deserves a
special investigation and we shall return to it in a separate paper
(see~\cite{bp-NF-D+EP}). Below we sketch an equivalent technique for
an electromagnetic field.

	Recall~\cite{L&L-2,Drechsler&Mayer}, the classical electromagnetic
field is described via a real 1\ndash form $A$ over a 4\ndash dimensional
real manifold $M$ (endowed with a (pseudo-)Riemannian metric $g$ and)
representing the space\ndash time model and, usually, identified with the
Minkowski space $M^4$ of special relativity or the (pseudo\ndash)Riemannian
space $V_4$ of general relativity.~%
\footnote{~%
The particular choice of $M$ is insignificant for the following.%
}
The electromagnetic field itself is represented by the two\ndash form
$F=\od A$, where ``$\od$'' denotes the exterior derivative operator, with
local components (in some local coordinates $\{x^\mu\}$)
	\begin{equation}	\label{4-5n.18}
F_{\mu\nu}
= - \frac{\pd A_\mu}{\pd x^\nu} +  \frac{\pd A_\nu}{\pd x^\mu} .
	\end{equation}

	As it is well known, the electromagnetic field, the
Maxwell equations describing it,
 and its (minimal) interactions with other objects are
invariant under a gauge transformation
	\begin{equation}	\label{4-5n.19}
A_\mu \mapsto A'_\mu
	= A_\mu + \frac{\pd \lambda}{\pd x^\mu}
	\end{equation}
or $A\mapsto A'=A+\od \lambda$, where $\lambda$ is a $C^2$ function. As is
almost evident, the electromagnetic field is invariant under simultaneous
changes of the local coordinate frame,
 $E_\mu=\frac{\pd}{\pd x^\mu}\mapsto E'_\mu=B_\mu^\nu E_\nu$
with $B_\mu^\nu:=\frac{\pd x^\nu}{\pd x^{\prime\,\mu}}$,
and a gauge transformation~\eref{4-5n.19}:
	\begin{equation}	\label{4-5n.20}
A_\mu \mapsto A'_\mu
	= B_\mu^\nu A_\nu + E'_\mu(\lambda)
	= B_\mu^\nu \Bigl(A_\nu + \frac{\pd \lambda}{\pd x^\nu} \Bigr) .
	\end{equation}
A simple calculation shows that under the transformation~\eref{4-5n.20}, the
quantities~\eref{4-5n.18} transform like components of an antisymmetric tensor,
	\begin{equation}	\label{4-5n.21}
F_{\mu\nu}\mapsto F'_{\mu\nu} = B_\mu^\sigma B_\nu^\tau F_{\sigma\tau}
	\end{equation}
due to which the 2-form $F$ remains unchanged, $F=\od A=\od A'$. Notice,
above $A'_\mu$ are \emph{not} the components of $A$ in $\{E'_\mu\}$ unless
$\lambda=\const$ while  $F_{\mu\nu}^{\prime}$ \emph{are} the components of
$F$ in $\big\{E^{\prime\mu}=\frac{\pd x^{\prime\mu}}{\pd x^\nu} \od
x^{\nu}\big\}$.

	The similarity between~\eref{4-5n.20} and~\eref{4-5n.12}
is obvious and implies the idea of
identifying (on an open set, neighborhood) the  electromagnetic potentials
$A_\mu$ with the matrices (functions, in the particular case) $\Gamma_\mu$ of
the 3\ndash index coefficients of some linear transport along paths in a
1\ndash dimensional vector bundle $(E,\pi,M)$. This can be done as follows.

	Let $M$ be a real 4-dimensional manifold, representing the
space\ndash time  model, and $(E,\pi,M)$  be a 1\ndash dimensional real
vector bundle over it.~%
\footnote{~%
The consideration of the real case does not change the above results with the
exception that $\mathbb{C}$ should be replaced by $\mathbb{R}$.%
}
We
\emph{%
identify the potentials $A_\mu$ of an electromagnetic field with the (local)
coefficients of a linear transport $L$ along paths in $(E,\pi,M)$ whose
matrix has the representation~\eref{4-5n.9} (along every path and in every
pair of frames)%
}.
Hence, the 3-index coefficients of $L$ are uniquely defined and supposed
to be (arbitrarily) fixed in some pair of frames.

	Since the 3-index coefficients of linear transport are defined in a
pair of frames $(\{e\},\{E_\mu\})$, $\{e\}$ in the bundle space $E$ and
$\{E_\mu\}$ in the tangent bundle space $T(M)$, the change~\eref{4-5n.20}
expresses simply the transformation of $A_\mu$ under the pair of changes
$e\mapsto e'=ae$ and $E_\mu\mapsto E'_\mu=B_\mu^\nu E_\nu$  and is a
consequence of~\eref{4-5n.12} if we put
	\begin{equation}	\label{4-5n.22}
a=\e^\lambda .
	\end{equation}
It should be emphasized, now the (pure) gauge transformation~\eref{4-5n.19}
appears as a special case of~\eref{4-5n.20}, corresponding to a change of
the frame in $E$ and a fixed frame in $T(M)$.~%
\footnote{~%
Cf.\ a similar conclusion in~\cite[p.~178]{Nash&Sen}, in which a gauge
transformation, in a general gauge theory, is interpreted as a change in
fibre coordinates of a principle bundle.%
}
This means that, in the approach proposed, the change~\eref{4-5n.19} is
directly incorporated in the definition of the field potential $A$. This
conclusion is in contrast to the situation in classical electrodynamics as
there the change~\eref{4-5n.19} is a simple observation of `additional'
invariance of the field, which is not connected with the geometrical
interpretation of the theory.

	Defining the electromagnetic field (strength) by $F=\od A$,
the equality~\eref{4-5n.18} remains valid in a coordinate frame
$\{E_\mu=\pd/\pd x^\mu\}$. Since $A$ and $F$ possess all of the properties
they must have in classical electrodynamics, they represent an equivalent
description of electromagnetic field. The only difference with respect to
the classical description is the clear geometrical meaning of these
quantities, as a consequence of which an electromagnetic field can be
identified with a linear transport along paths in a one\ndash dimensional
vector bundle over the space\ndash time.
	With a little effort, one can show that the proposed treatment of
electromagnetic field is equivalent to the modern one in the bundle picture
of gauge theories (see, e.g.,~\cite{Konopleva&Popov} or~\cite{Baez&Muniain}),
where the electromagnetic potentials are regarded as coefficients of a
suitable linear connection.

	In the approach proposed, the different gauge conditions, which are
frequently used, find a natural interpretation as a partial fix of the class
of frames in the bundle space employed. For instance, any one of the gauges
in the table~\ref{4-TableOfGauges}~\vpageref{4-TableOfGauges} corresponds to a
class of frames for which~\eref{4-5n.20} holds for $B_\mu^\nu=\delta_\mu^\nu$,
$\delta_\mu^\nu$ being the Kronecker deltas, and $\lambda$ subjected to a
condition given in the table.~%

\footnote{~%
Below $M$ is supposed  to be endowed with a (pseudo-)Riemannian metric $g_{\mu\nu}$,
the coordinates to be numbered as $x^0,x^1,x^2$, and $x^3$, $x^0$ to be the
`time' coordinate, $\pd_\mu:=\pd/\pd x^\mu$, and $\pd^\mu:=g^{\mu\nu}\pd_\nu$
with $[g^{\mu\nu}]:=[g_{\mu\nu}]^{-1}$.%
}

\vspace{-2.3ex}
	\begin{table}[ht!]
\caption{Examples of gauge conditions \label{4-TableOfGauges}}
\vspace{1ex}
	\begin{minipage}{\textwidth}
	\begin{tabularx}{\textwidth}{lrll@{}}
Gauge  & Condition on $A$  & Condition on $\lambda$  & Condition on $\varphi$
\\ \hline
Lorenz~%
\footnote{~%
The Lorenz condition and gauge
are named in honor of the Danish theoretical physicist
Ludwig Valentin Lorenz (1829--1891),
who has first published it in 1867~\cite{Lorenz/1867}
(see also~\cite[pp.\ 268-269, 291]{Whittaker-History});
however this condition was first introduced in lectures by Bernhard
G.~W.~Riemann in 1861 as pointed in~\cite[p.~291]{Whittaker-History}. It
should be noted that the \emph{Lorenz} condition/gauge is quite often
erroneously referred to as the Loren\emph{t}z condition/gauge after the name of
the Dutch theoretical physicist Hendrik Antoon Lorentz (1853--1928) as, e.g.,
in~\cite[p.~18]{Roman-QFT} and in~\cite[p.~45]{Gockeler&Schucker}.%
}
	& $\pd^\mu A_\mu =0$
	& $\pd^\mu\pd_\mu \lambda =0$
	& $\pd^\mu\pd_\mu \varphi = - \pd^\mu\pd_\mu \lambda $
\\
Coulomb\footnote{In this row the summation over $k$ is from 1 to 3.}
	& $\pd^k A_k =0$
	& $\pd^k \pd_k \lambda =0$
	& $\pd^k \pd_k \varphi
		=- \pd^k \pd_k \lambda $
\\
Hamilton
	& $A_0=0$
	& $\lambda(x) = \lambda(x^1,x^2,x^3)$
	& $\varphi(x) = \varphi(x^1,x^2,x^3)$
\\
Axial
	& $A_3=0$
	& $\lambda(x) = \lambda(x^0,x^1,x^2)$
	& $\varphi(x) = \varphi(x^0,x^1,x^2)$
\\ \hline
	\end{tabularx}
	\end{minipage}
	\end{table}
In the table~\ref{4-TableOfGauges}~\vpageref{4-TableOfGauges} $\varphi$ is a
$C^1$ function describing the arbitrariness in the choice of $\lambda$, \ie if
a gauge condition is valid for $\lambda$, then it holds also for
$\lambda+\varphi$ instead of $\lambda$.

\section{Normal and inertial frames}
	\label{Sect6}

	Comparing~\eref{4-5n.18} with~\eref{4-5.6}, we get~%
\footnote{~%
In this section, we assume the Greek indices to run over the range
0, 1, 2, 3.%
}
	\begin{equation}	\label{4-5n.23}
F_{\mu\nu} = R_{\mu\nu}(-A_0,-A_1,-A_2,-A_3) .
	\end{equation}
Thus, the electromagnetic field tensor  $F$ is completely responsible for the
existence of frames normal for $L$ (theorems~\ref{4-Thm5.1}
and~\ref{4-Thm5.2}). For example, if $U$ is an open set, frames normal on
$U\subseteq M$ for $L$ exist iff $F|_U=0$, \ie if electromagnetic field
is missing on $U$.~%
\footnote{~%
Elsewhere we shall prove that the components $F_{\mu\nu}$ completely describe
the curvature of $L$ which agrees with the interpretation of $F_{\mu\nu}$ as
components of the curvature of a connection on a vector bundle in the gauge
theories~\cite{Konopleva&Popov,Drechsler&Mayer,Slavnov&Fadeev}. The general
situation is similar: the quantities~\eref{4-5.6} determine the curvature of
a transport with coefficients' matrix~\eref{4-5.1}.%
}
Also, if $N$ is a submanifold of $M$, frames normal on $U$ for $L$ exist iff
in the special coordinates $\{x^\mu\}$, described in theorem~\ref{4-Thm5.2},
is valid $F_{\alpha\beta}|_U=0$ for  $\alpha,\beta=1,\dots,\dim N$.
In the context of theorem~\ref{4-Thm3.1}, we can say
that an electromagnetic field admits frames normal on $U\subseteq M$ iff
the linear transport $L$ corresponding to it is path\ndash independent on
$U$ (along paths lying entirely in $U$). Thus, if $L$ is path\ndash dependent
on $U$, the field does not admit frames normal on $U$. This important result
is the classical analogue of a quantum effect, know as the Aharonov\ndash
Bohm effect~\cite{Aharonov&Bohm, Bernstein&Phillips}, whose essence is that
the electromagnetic potentials directly, not only through the field tensor
$F$, can give rise to observable physical results.

	Let us now turn our attention to the physical meaning of the normal
frames corresponding to a given electromagnetic field which is described, as
pointed above, via a linear transport $L$ along paths in a line vector bundle
over the space\ndash time $M$.

	Suppose $L$ is Euclidean on a neighborhood $U\subseteq M$. As a
consequence of~\eref{4-5n.23} and theorem~\ref{4-Thm5.1}, we have $F|_U=\od
A|_U=0$, \ie on $U$ the electromagnetic field strength vanishes and hence the
field is a pure gauge on $U$,
	\begin{equation}	\label{4-5n.24}
A_\mu|_U = \frac{\pd f_0}{\pd x^\mu}\Big|_U
	\end{equation}
for some  $C^1$ function $f_0$ defined on an open set containing $U$ or equal
to it. As we know from proposition~\ref{4-Prop3.1}, in a frame $\{e'\}$ normal
on $U$ for $L$ vanish the 2\ndash index coefficients of $L$ along any path
$\gamma$ in $U$:
	\begin{equation}	\label{4-5n.25}
\Mat{\Gamma}'(s;\gamma) = A'_\mu(\gamma(s)) \dot\gamma^\mu(s) = 0
	\end{equation}
for every $\gamma\colon J\to U$ and $s\in J$. Using~\eref{4-5n.24}, it is
trivial to see that any transformation~\eref{4-5n.20} with
	\begin{equation}	\label{4-5n.26}
\lambda = -f_0
	\end{equation}
transforms $A_\mu$ into $A'_\mu$ such that
	\begin{equation}	\label{4-5n.27}
A'_\mu|_U = 0
	\end{equation}
(irrespectively of the frames $\{E_\mu\}$ and $\{E'_\mu\}$ in the tangent
bundle over $M$). Hence, by~\eref{4-5n.25} the one\ndash vector frame
$\{e'=\e^{-f_0}e\}$ in the bundle space $E$ is normal for $L$ on $U$.
Therefore, in the frame $\{e'\}$, vanish not only the 2\ndash index
coefficients of $L$ but also its \emph{3\ndash index} ones, \ie $\{e'\}$ is a
frame \emph{strong normal} on $U$ for $L$. Applying~\eref{4-5n.20} one can
verify,
\emph{%
all frames strong normal on a neighborhood $U$ for $L$ are obtainable from
$\{e'\}$ by multiplying its vector $e'$ by a function $f$ such that
$\frac{\pd f}{\pd x^\mu}\bigr|_U=0$%
},
\ie they are $\{b\e^{-f_0}e\}$ with $b\in\mathbb{R}\backslash\{0\}$ as  $U$ is
a neighborhood. Thus, every frame normal on a neighborhood $U$ for $L$ is
strong normal on U for $L$ and vice versa.

	A frame (of reference) in the bundle space in which~\eref{4-5n.27}
holds on a subset $U\subseteq M$, will be called \emph{inertial on $U$ for the
electromagnetic field} considered. In other words, the frames inertial on $U$
for a given electromagnetic field are the ones in which its potentials vanish
on $U$. Thus, every frame inertial on $U$ is strong normal on it and vice
versa.

	So, in a frame inertial on $U\subseteq M$ for an electromagnetic
field it is not only a pure gauge, but in such a frame its potentials vanish
on $U$.  Relying on the results obtained
(see~\cite{bp-Frames-n+point,bp-Frames-path,bp-Frames-general}), we can assert
the existence of frames inertial at a single point and/or along paths without
self\ndash intersections for every electromagnetic field, while on
submanifolds of dimension not less than two such frames exist only as an
exception if (and only if) some additional conditions are satisfied, \ie for
some particular types of electromagnetic fields.

	Now we would like to make a link with the paper~\cite{bp-PE-P?} in
which was demonstrated that the (ordinary strong) equivalence principle is a
provable theorem and the inertial frames in a gravity theory based on a
linear connection (or other derivation) are the frames normal for it. For
reasons given a few lines below, such frames will be called \emph{inertial
for the gravitational field} under consideration.

	Let there be given a physical system consisting of pure
or, possibly, interacting gravitational and electromagnetic fields which are
described via, respectively, a linear connection $\nabla$ in the tangent
bundle $(T(M),\pi_T,M)$ (or the tensor algebra) over the space\ndash time
$M$ and a linear transport along paths in a 1\ndash dimensional vector bundle
$(E,\pi_E,M)$ over $M$. On one hand, as we saw above, the frames inertial
for an electromagnetic field, if any, in the bundle space $E$ are completely
independent of any frame in the bundle space $T(M)$ tangent to $M$. On the
other hand, the frames inertial for the gravity field, \ie the ones normal
for $\nabla$, if any, are frames in $T(M)$ and have nothing in common with the
frames in $E$, in particular with the frames normal for $L$, if any.
Consequently, if there is a frame $\{E_\mu\}$ in $T(M)$ inertial on
$U\subseteq M$ for the gravity field and a frame $\{e\}$ in $E$ inertial on
the same set $U$ for the electromagnetic field, the frame
$\{e\times E_\mu\}=\{(e,E_\mu)\}$ in the bundle space of the bundle
$(E\times T(M),\pi_E\times\pi_T,M\times M)$
over $M\times M$ can be called
simply \emph{inertial} on $U$ (for the system of gravity and electromagnetic
fields).~%
\footnote{~%
For purposes which will be explained elsewhere, the product bundle
$(E\times T(M),\pi_E\times\pi_T,M\times M)$ is better to be replaced with the
bundle $(\mathsf{F},\pi,M)$, with
 $\mathsf{F}:=\{ (\xi,\eta)\in E\times T(M) | \pi_E(\xi)=\pi_T(\eta) \}$
and $\pi(\xi,\eta):=\pi_E(\xi)=\pi_T(\eta)\in M$ for
$(\xi,\eta)\in\mathsf{F}$,
\ie $\pi^{-1}(x)= \pi_E^{-1}(x) \times \pi_T^{-1}(x)$, $x\in M$. Evidently,
$(\mathsf{F},\pi,M)$ is isomorphic to the Whitney sum~\cite[sect.~1.29]{Poor}
of $(E,\pi_E,M)$ and $(T(M),\pi_T,M)$ and the standard fibre of
$(\mathsf{F},\pi,M)$ can be identified with
$\mathbb{R}\times\mathbb{R}^4=\mathbb{R}^5$ as $(E,\pi_E,M)$ is 1\ndash
dimensional and  $\dim M=4$.%
}
Thus, in an inertial frame, if any, the potentials of both, gravity
and electromagnetic, fields vanish. Relying on the results obtained in this
work, as well as on the ones
in~\cite{bp-Frames-n+point,bp-Frames-path,bp-Frames-general,bp-PE-P?}, we can
assert the existence of inertial frames at every single space\ndash time
point and/or along every path without self\ndash intersections in it. On
submanifolds of dimension higher than one, inertial frames exist only for
some exceptional configurations of the fields which can be described on
the base of the results in the cited works.

\section{Conclusion}
    \label{Conclusion}

	One of the purposes of the present paper was to exemplify the general
theory of linear transports along paths and the frames normal for them on line
bundles. The application of the so\ndash obtained results to the classical
electrodynamics gives rise to a geometric interpretation of the electromagnetic
field as a linear transport in a line bundle and to an introduction of
inertial frames for this field. As pointed in~\cite{bp-EPinED}, the linear
transport, describing the electromagnetic field in our approach, is in fact the
parallel transport generated by the linear connection describing it in the well
known its geometrical interpretation~\cite{Gockeler&Schucker}.

	The coincidence of the normal and inertial frames for the
electromagnetic field expresses the equivalence principle for that
field~\cite{bp-EPinED}. Generally this principle is a provable theorem and it
is always valid at any single point of along given path (without
selfintersections) as these are the only cases when normal frames for a linear
connection or transport always exist.

	For a free electromagnetic field, the line bundle mentioned above
remains unspecified. However, if an interaction of that field with other one is
presented, the line bundle under question can be identified or uniquely
connected with a bundle (over the spacetime) whose sections represent the
latter field. Moreover, in such a situation the equivalence principle can be
used to justify the so\ndash called minimal coupling (principle).

	The considerations in this work confirm our opinion that the frames
(and possibly coordinates) in bundle spaces, in which some physical fields
`live', should be regarded as parts of the frames of references with respect to
which a physical system is investigated.~%
\footnote{~%
Cf.~\cite[sect.~1]{Fatibene&et._al.-2004} where similar ideas relative to
conventions concerning physical laboratories can be found.
}
In the particular case of an electromagnetic field, these are the one\ndash
vector field frames $\{e\}$ in the bundle space $E$ of the line bundle
$(E,\pi,M)$ in which the field is describe via a linear transport $L$ along
paths. With respect to $\{e\}$ is defined the sole coefficient of $L$ and with
respect to a pair $(\{e\},\{E_\mu\})$, with $\{E_\mu\}$ being a frame in the
bundle tangent to the spacetime, are defined the (3\ndash index) coefficients
of $L$ which, by definition, coincide with the components  $A_\mu$ of the
4\ndash vector potential of the electromagnetic field. Since $A_\mu$ are
observable (if one beliefs in the Aharonov-Bohm effect) and $\{E_\mu\}$, usually
constructed from some local coordinates
$x^\mu$ ($E_\mu=\frac{\pd }{\pd x^\mu}$), is an essential path of the frames of
reference, one can conclude that $\{e\}$ should be a part of the frame of
reference and there should exist a method of its experimental/ laboratory
realization.


\section*{Acknowledgments}

	The work on this paper was partially supported by the National Science
Fund of Bulgaria under Grant No.~F~1515/2005.


\addcontentsline{toc}{section}{References}
\bibliography{bozhopub,bozhoref}
\bibliographystyle{unsrt}
 \addcontentsline{toc}{subsubsection}{This article ends at page}

\end{document}